\documentclass{article}
\usepackage{spconf,amsmath,graphicx}

\title{Channel covariance estimation in multiuser massive MIMO systems with an approach based on infinite dimensional Hilbert spaces}

\name{Renato~L.~G.~Cavalcante and S.~Sta\'nczak}
\address{Fraunhofer Heinrich Hertz Institute, Berlin, Germany \\ email: $\{$renato.cavalcante,slawomir.stanczak$\}$@hhi.fraunhofer.de}
\usepackage{ifpdf}
\usepackage{manfnt}
\usepackage[mathscr]{eucal}

\usepackage{graphicx}

\usepackage{algorithmicx}
\usepackage[ruled,vlined,commentsnumbered]{algorithm2e}
\usepackage{amssymb}

\usepackage{amsthm}

\usepackage{caption}
\usepackage{subcaption}
\usepackage{cite}
\usepackage{color}
\usepackage{colortbl}
\definecolor{colorref}{rgb}{0.4648,0,0} 
\definecolor{colorcite}{rgb}{0,0.2902,0.1765}

\usepackage{amsmath}
\newtheorem{proposition}{Proposition}

\newtheorem{Cor}{Corollary}

\newcommand{\signal}[1]{{\boldsymbol{#1}}}
\newcommand{\real}{{\mathbb R}}
\newcommand{\innerprod}[2]{\left\langle{#1},{#2}\right\rangle}
\newcommand{\innerprodh}[3]{\left\langle{#1},{#2}\right\rangle_{#3}}

\newtheorem{example}{Example}
\newcommand{\Natural}{{\mathbb N}}
\newcommand{\refeq}[1]{(\ref{#1})}

\newcommand{\dprime}{{\prime\prime}}
\newcommand{\ha}{{\mathcal{H}_1}}
\newcommand{\hb}{{\mathcal{H}_2}}
\newcommand{\hc}{{\mathcal{H}_3}}

\begin{document}
\ninept

\maketitle

\begin{abstract}
We propose a novel algorithm to estimate the channel covariance matrix of a desired user in multiuser massive MIMO systems. The algorithm uses only knowledge of the array response and rough knowledge of the angular support of the incoming signals, which are assumed to be separated in a well-defined sense. To derive the algorithm, we study interference patterns with realistic models that treat signals as continuous functions in infinite dimensional Hilbert spaces. By doing so, we can avoid common and unnatural simplifications such as the presence of discrete signals, ideal isotropic antennas, and infinitely large antenna arrays. An additional advantage of the proposed algorithm is its computational simplicity: it only requires a single matrix-vector multiplication. In some scenarios, simulations show that the estimates obtained with the proposed algorithm are close to those obtained with standard estimation techniques operating in interference-free and noiseless systems.
\end{abstract}
\section{Introduction}
\label{sec:intro}

Estimation of channel covariance matrices is a key ingredient of many receiver architectures proposed for massive MIMO systems \cite{miretti18,miretti18SPAWC,renato18error,haghighatshoar2017massive,bjornson16,dec2015,dai2018,emil18}. For example, in multiuser systems, knowledge of channel covariance matrices can be exploited to devise efficient spatial multiplexing schemes \cite{ad2013}. However, estimating these matrices is not trivial because of, for example, the problems caused by pilot contamination. In this study, we address this estimation task. We propose a novel technique to separate the channel covariance matrix of a user of interest from the covariance matrix of the received signal, which is assumed to be contaminated with interference and noise.  

In more detail, we start by investigating the influence of the antenna array response on the interference between users in massive MIMO systems. If the number of antennas is finite, we show an alternative way of proving that two users can cause strong mutual interference even if there is no overlap in the angle-of-arrival of the signals. This result can be used, for example, in the derivation of location-aided schedulers that avoid many simplifications in the literature, but we do not consider these schedulers in detail here because they are not the focus of this study. In the proposed analysis of interference patterns, we show that, from a mathematical perspective, antenna arrays can be seen as bounded linear operators that map infinite-dimensional Hilbert spaces representing incoming signals to finite-dimensional Hilbert spaces representing channel covariance matrices. This connection with operator theory enables us to derive a simple algorithm to separate covariance matrices of interfering users. The algorithm uses only knowledge of the array response and coarse knowledge of the angles of arrival of the interfering signals, which are assumed to be separated in the angular domain in a well-defined sense. Furthermore, the algorithm is based on simple linear operations, and simulations show that its performance can be similar to that obtained with sample-based estimation techniques operating in interference-free and noiseless massive MIMO systems. 

The results we derive are related to those in \cite{dai2018,mupp18,zhao2015,yin2014,xie2018channel,haghighatshoar2017massive}, but the proposed framework is more general in some aspects. For example, it can be applied to realistic propagation models (e.g., models considering cross-polarized antenna arrays \cite{miretti18SPAWC}) without any changes. Furthermore, widely-used simplifications and approximations, such as the discretization of the angular power spectra, uniform linear arrays, ideal isotropic antennas, and infinitely large arrays are not required. (However, for simplicity and because of the space limitation, we restrict the \emph{numerical examples} -- but not the theory and the algorithmic framework -- to uniform linear arrays.) Moreover, contrasting with previous studies, we use infinite dimensional Hilbert spaces to decouple the effects of the angular power spectra and the array response on interference patterns. We remark that these  spaces have been used in the context of massive MIMO systems in  \cite{miretti18,miretti18SPAWC,renato18error}, but these existing studies have not considered interference in multiuser environments. Algorithms for covariance estimation in multiuser massive MIMO have been previously proposed in, for example, \cite{upadhya2018covariance,neumann2018covariance,bjornson2016massive}, but the algorithms in those studies typically require specific training pilots that we do not assume to be available.

We omit all proofs in this study because of the space limitation.

\section{Preliminaries}
\label{sec:math}

Hereafter, the operators $(\cdot)^t$ and $(\cdot)^H$ denote, respectively, the transpose and the Hermitian transpose of matrices or vectors. The set $\real_+$ is the set of nonnegative reals, and $i$ is the imaginary number; i.e., the solution to $i^2=-1$. The real and imaginary components of a complex matrix $\signal{M}\in\mathbb{C}^{N\times N}$ are given by, respectively, $\mathrm{Re}(\signal{M})\in\real^{N\times N}$ and $\mathrm{Im}(\signal{M})\in\real^{N\times N}$. 

Given two real Hilbert spaces $\left(\mathcal{H}^\prime, \innerprod{\cdot}{\cdot}_{\mathcal{H}^\prime}\right)$ and $\left(\mathcal{H}^\dprime,\innerprod{\cdot}{\cdot}_{\mathcal{H}^\dprime}\right)$, the set $\mathcal{B}(\mathcal{H}^\prime, \mathcal{H}^\dprime)$ is the set of bounded linear operators that map $\mathcal{H}^\prime$ to $\mathcal{H}^\dprime$. The adjoint of $T\in \mathcal{B}(\mathcal{H}^\prime, \mathcal{H}^\dprime)$ is denoted by $T^*\in \mathcal{B}(\mathcal{H}^\dprime, \mathcal{H}^\prime)$, and it is defined by the equality $(\forall x\in\mathcal{H}^\prime)(\forall y\in\mathcal{H}^\dprime)\innerprod{Tx}{y}_{\mathcal{H}^\dprime} = \innerprod{x}{T^* y}_{\mathcal{H}^\prime}$. 

 If $\Omega$ is a given set and $S\subset\Omega$, the  function $1_S:\Omega\to\{0,1\}$ takes the value $1_S(\theta)=1$ if $\theta\in S$ or the value $1_S(\theta)=0$ otherwise. The set $2^\Omega$ is the powerset of $\Omega$. For two functions $f_1:\Omega\to\real$ and $f_2:\Omega\to\real$, we define $f_1\cdot f_2:\Omega\to\real:x\mapsto f_1(x)f_2(x)$. The set $\mathrm{Supp} f=\{x \in \Omega~|~f(x) \neq 0\}$ is the support of $f:\Omega\to\real$, and we define $\mathrm{ess~sup}_{x\in\Omega}~ f(x) := \inf\{c\in\real~|~\mu(\{x\in\Omega~|~f(x)>c\})=0\}$ for measurable functions $f$  w.r.t. the measure $\mu$.  By $\mathcal{N}_\mathbb{C}(\signal{m}, \signal{R})$ we denote the distribution of a complex multivariate normal distribution with mean $\signal{m}$ and covariance $\signal{R}$.  

For convenience, we use a \emph{real} vector space with vectors being complex matrices in $\mathcal{H}_1:=\mathbb{C}^{N\times N}$. In this real vector space, matrix sum is defined in the usual way, and scalar multiplication is defined only for real scalars. Equipping $\mathcal{H}_1$ with the inner product 
$(\forall \signal{M}_1\in\mathcal{H}_1)(\forall \signal{M}_2\in\mathcal{H}_1) \innerprodh{\signal{M}_1}{\signal{M}_2}{\ha} := \mathrm{Re}(\mathrm{tr}(\signal{M}_2^H\signal{M}_1))$,
we obtain the \emph{real} Hilbert space denoted by $(\mathcal{H}_1, \innerprodh{\cdot}{\cdot}{\ha})$. This inner product can be equivalently computed by $
(\forall \signal{M}_1\in\mathcal{H}_1)(\forall \signal{M}_2\in\mathcal{H}_1) \innerprodh{\signal{M}_1}{\signal{M}_2}{\ha} = \innerprodh{T_\mathrm{vec}(\signal{M}_1)}{T_\mathrm{vec}(\signal{M}_2)}{\hb},$
where 
\begin{align}
\label{eq.vec_op}
T_\mathrm{vec}:\mathcal{H}_1\to\mathcal{H}_2: \signal{M}\mapsto \mathrm{vec}\left(\left[\begin{matrix}\mathrm{Re}(\signal{M})~ \mathrm{Im}(\signal{M})\end{matrix}\right]\right)
\end{align} 
is the operator that vectorizes a matrix and $(\mathcal{H}_2,\innerprodh{\cdot}{\cdot}{\hb})$ is the real Hilbert space with $\mathcal{H}_2:=\real^{2N^2}$ and $(\forall \signal{x}\in\mathcal{H}_2)(\forall \signal{y}\in\mathcal{H}_2)~\linebreak[4] \innerprodh{\signal{x}}{\signal{y}}{\hb}:=\signal{x}^t\signal{y}.$ 

\section{System Model}

\label{sect:model}

In the uplink of a system with $M\in\Natural$ single-antenna users sending pilots to a base station equipped with $N\in\Natural$ antennas, we denote by $\mathcal{U}=\{1,\ldots, M\}$ the set of users and by $\signal{h}_j(t)\in\mathbb{C}^N$ the channel sample of user $j$ at time $t\in\real$. We use the conventional ergodic wide-sense stationary (WSS) assumption \cite{miretti18,emil18,yin2013,yin2014,ad2013}, and the $k$th channel sample $\signal{h}_j[k]:=\signal{h}_j(k T_\mathrm{c})$ ($k\in\Natural$) of user $j$ is drawn from the distribution  $\mathcal{N}_\mathbb{C}(\signal{0},\signal{R}_j)$, where $T_\mathrm{c}$ is the channel coherence time and $(\forall k\in\Natural) E[\signal{h}_j[k]\signal{h}_j[k]^H]=\signal{R}_j\in\mathcal{H}_1$ is the channel covariance matrix.\footnote{To avoid notational clutter, we  use the same notation for random variables and their samples. The definition that should be applied is clear from the context.}

Let $\signal{U}_j \signal{S}_j\signal{U}^H_j = \signal{R}_j$ denote the  eigendecomposition of the channel covariance matrix of user $j$, where $\signal{U}_j\in\mathbb{C}^{N\times N}$ is a unitary matrix and $\signal{S}_j\in\real_+^{N\times N}$ is a diagonal matrix. As common in the literature \cite{ad2013,dec2015}, the channel sample $\signal{h}_j[k]$ at time $k\in\Natural$ of a given user $j$ is further assumed to take the form $\signal{h}_j[k]=\signal{U}_j \signal{S}_j^{1/2}\signal{w}_j[k]$, where $(\signal{w}_j[k])_{k\in\Natural}\subset\mathbb{C}^N$ are samples of i.i.d. random vectors with distribution $\mathcal{N}_\mathbb{C}(\signal{0},\signal{I})$, and  $(\forall k\in\Natural)~ (\signal{w}_j[k])_{j\in\mathcal{U}}$ are samples of mutually independent random vectors.

With the above assumptions, at time $k\in\Natural$, the uplink signal received at the base station spaced by multiples of the coherence interval $T_c$ in a memoryless (block) flat fadding channel is given by

\begin{align}
\label{eq.signal}
\signal{y}[k] = \sum_{j\in\mathcal{U}} \signal{h}_j[k]~s_j[k] + \signal{n}[k] \in \mathbb{C}^N,
\end{align}
where $s_j[k]\in\mathbb{C}$ and $\signal{h}_j[k]\in\mathbb{C}^N$ denote, respectively, the pilot symbol and the channel of user $j$, and $\signal{n}[k]\in\mathbb{C}^N$ is the noise sample. In this study, we impose few assumptions on the pilots of different users (they can even be the same). In particular, for a given user $j$, we only assume that the pilots are independent from channels and they have bounded second moments. Without any loss of generality, the pilots $(s_j[k])_{k\in\Natural}$ satisfy $(\forall j\in\mathcal{U})(\forall k\in\Natural)~E[|s_j(k)|^2]=1$. Noise samples $(\signal{n}[k])_{k\in\Natural}$ are drawn from i.i.d. random vectors distributed according to  $\mathcal{N}_\mathbb{C}(\signal{0},~\sigma^2\signal{I})$ with $\sigma>0$, so we have

\begin{align}
\label{eq.contamination}
(\forall k\in\Natural)~E[\signal{y}[k]\signal{y}[k]^H] = \signal{R} = \sum_{j\in\mathcal{U}} \signal{R}_j + \sigma^2 \signal{I}.
\end{align}

The distributions of $\signal{y}[k]$ and the channels $(\signal{h}_j[k])_{j\in\mathcal{U}}$ do not depend on $k\in\Natural$ (because of the WSS assumption), so in the text that follows we drop the time index $k$ to simplify the notation of the random variables if confusion does not arise.

\section{Favorable propagation and pilot decontamination}
\label{sect.proposed}
As discussed in \cite[Ch.~7]{marzetta2016}, an important task in massive MIMO systems is to identify whether the channels of two distinct users $j$ and $l$ are orthogonal or approximately orthogonal, or, more precisely, identify whether $|\signal{h}_j^H\signal{h}_l|$ is zero or close to zero with high probability. In these cases, we say that the channels offer \emph{favorable propagation} or \emph{approximately favorable propagation} \cite[p. 139]{marzetta2016}. In Sect.~\ref{sect:favorable_propagation}, we derive results that enable us to perform this identification task from coarse knowledge of the interfering angular power spectra and the array response.  Then, in Sect.~\ref{sect:algorithms}, we introduce a simple algorithm for channel covariance estimation in multiuser systems. All these results require connections between the angular power spectra, the channel covariance matrices, and the antenna array response in an infinite dimensional Hilbert space. These connections are established in the next subsection.

\subsection{Relation between the angular power spectrum, the array response, and the channel covariance matrix}
\label{sect.relations}

As shown in \cite{miretti18,miretti18SPAWC,renato18error} (see also Example~\ref{example.ULA} below), a common feature of massive MIMO systems considering realistic antenna and propagation models is the fact that the $n$th component ($1\le n\le 2N^2$) of the vector $\signal{r}_j:= T_\mathrm{vec}(\signal{R}_j)\in\mathcal{H}_2$ of user $j\in\mathcal{U}$ is given by
\begin{align}
\label{eq.innerprod}
r_{j,n} = \innerprodh{\rho_j}{g_n}{\hc},
\end{align}
where $(\rho_j, g_n)\in\mathcal{H}_3\times\mathcal{H}_3$, $(\mathcal{H}_3,\innerprodh{\cdot}{\cdot}{\hc})$ is a real Hilbert space (of equivalent classes) of squared-integrable functions $\mathcal{H}_3=L^2(\Omega)$ equipped with the inner product $(\forall x\in\mathcal{H}_3) (\forall y\in\mathcal{H}_3)~\innerprodh{x}{y}{\hc}=\int_{\Omega} x~y~\mathrm{d}\mu$, the set $\Omega\subset \real^K$ is a compact set used to represent altitude and azimuth angles, and $\mu$ is the standard Lebesgue measure on $\real^K$. The function $\rho_j:\Omega\to\real_+$ is the so called \emph{angular power spectrum} of user $j$, and it is a member of the cone $\mathcal{K}:=\{\rho\in\mathcal{H}_3~|~\mu(\{\theta\in\Omega~|~\rho(\theta)< 0\}) = 0 \}$ of $\mu$-almost everywhere (a.e.) nonnegative functions. The functions $(g_n)_{n\in\{1,\ldots,2N^2\}}$ in $\mathcal{H}_3$ are the angular response of the array. 

We verify from \refeq{eq.innerprod} that arrays can be seen as bounded linear operators $T\in\mathcal{B}(\mathcal{H}_3,~\mathcal{H}_2)$ defined by

\begin{align}
\label{eq.bounded_op}
\begin{array}{rcl}
T:\mathcal{H}_3 &\to & \mathcal{H}_2 \\
\rho&\mapsto & \left[ \begin{matrix} \innerprodh{\rho}{g_1}{\hc}, \ldots,  \innerprodh{\rho}{g_{2N^2}}{\hc} \end{matrix} \right].
\end{array}
\end{align}

In the results that follow, the self-adjoint operator \linebreak[4] $T^*T:\mathcal{H}_3\to \mathcal{H}_3:\rho\mapsto \sum_{n=1}^{2N^2}\innerprodh{\rho}{g_n}{\hc} g_n $ plays a crucial role, and the next proposition shows additional properties that are required later. 

\begin{proposition}
	\label{prop.nonnegative}
	The self-adjoint operator $T^*T\in\mathcal{B}(\mathcal{H}_3,\mathcal{H}_3)$ maps (a.e.) nonnegative functions to (a.e) nonnegative functions, or, formally, $(\forall \rho\in\mathcal{K})~  T^*T(\rho)\in\mathcal{K}$. Furthermore, $T^*T$ is order-preserving with respect to the partial ordering induced by the cone $\mathcal{K}$; i.e., $(\forall \rho_1\in\mathcal{K})(\forall \rho_2\in\mathcal{K})~ \rho_1 \le_\mathcal{K} \rho_2 \Rightarrow  T^\star T(\rho_1) \le_\mathcal{K} T^* T(\rho_2)$, where the inequality $\le_\mathcal{K}$ is defined by $(\forall f_1 \in \mathcal{K})(\forall f_2 \in \mathcal{K}) f_1\le_\mathcal{K} f_2 \Leftrightarrow f_2 - f_1 \in\mathcal{K}$.
\end{proposition}

For concreteness, we exemplify the above concepts with an antenna array model that is widely used in the literature.

\begin{example} \label{example.ULA} Suppose that a base station is equipped with a uniform linear array (ULA), and it scans signals within the angular range $\Omega:=[-\pi/2,~\pi/2]$. In this case, \refeq{eq.innerprod} takes the form \cite{miretti18,miretti18SPAWC,haghighatshoar2017massive,xie2016overview,xie2018channel}
\begin{align*}
(\forall j\in\mathcal{U}) (\forall n\in\{1,\ldots,2N^2\})~r_{j,n}=\int_{-\pi/2}^{\pi/2} \rho_j(\theta){g}_n(\theta) \mathrm{d}\theta,
\end{align*} 
where $\rho_j(\theta)\in\real_+$ is the average  power from user $j\in\mathcal{U}$ arriving at the array at angle $\theta\in\Omega$. For ULAs, the functions $(g_n:\Omega\to\real)_{n\in\{1,\ldots,2N^2\}}$ in $\hc$ are given by 
\begin{align}
\label{eq.g}
(\forall\theta\in\Omega)~[g_1(\theta),\dots,g_{2N^2}(\theta)]^t = T_\mathrm{vec}(\signal{a}(\theta)\signal{a}(\theta)^H),
\end{align}
where 
\begin{align*}
\begin{array}{rl}
\signal{a}:\Omega\to\mathbb{C}^N:
\theta\mapsto \dfrac{1}{\sqrt{N}}\left[1,e^{i2\pi \frac{f}{c}d\sin\theta},\ldots,e^{i2\pi \frac{f}{c}d (N-1)\sin\theta}\right],
\end{array}
\end{align*}
is the array manifold, 
 $f$ is the operating frequency, $c$ is the speed of wave propagation,  and $d$ is the inter-antenna spacing. Therefore, the mapping $T\in\mathcal{B}(\mathcal{H}_3,\mathcal{H}_2)$ in \refeq{eq.bounded_op} reduces to
\begin{equation}
\label{eq.operator}
\begin{array}{rl}
T:\mathcal{H}_3\to&\mathcal{H}_2 \\
\rho\mapsto &\left[\int_{-\pi/2}^{\pi/2} \rho(\theta){g}_1(\theta) \mathrm{d}\theta,\ldots, \int_{-\pi/2}^{\pi/2} \rho(\theta){g}_{2N^2}(\theta) \mathrm{d}\theta\right].
\end{array}
\end{equation}
The adjoint $T^*\in\mathcal{B}(\mathcal{H}_2,\mathcal{H}_3)$ of $T$ is easily computed with the functions $(g_n)_{n\in\{1,\ldots,2N^2\}}$ in \refeq{eq.g}:
\begin{align}
\label{eq.adjoint}
T^*:\mathcal{H}_2\to\mathcal{H}_3:\left(x_1,\ldots,x_{2N^2}\right)\mapsto \sum_{n=1}^{2N^2}{x}_n g_n.
\end{align}
In addition, we deduce from \refeq{eq.operator} and \refeq{eq.adjoint} that $T^*T\in\mathcal{B}(\mathcal{H}_3,\mathcal{H}_3)$ maps $\rho\in\mathcal{H}_3$ to the function given by $(\forall\theta^\prime\in\Omega)$ 
\begin{align} 
\label{eq.smoothing}
(T^*T (\rho))(\theta^\prime)= \sum_{n=1}^{2N^2} \innerprodh{\rho}{g_n}{\hc} g_n(\theta^\prime)=\int_{-\pi/2}^{\pi/2} \kappa(\theta,\theta^\prime)\rho(\theta)\mathrm{d}\theta,
\end{align}
where $\kappa:\Omega\times\Omega\to\real:(\theta_1, \theta_2)\mapsto \sum_{n=1}^{2N^2} {g}_n(\theta_1){g}_n(\theta_2)$ is the kernel of the array. 
\end{example}

\subsection{The influence of the angular power spectra and the antenna array response on the interference between users}
\label{sect:favorable_propagation}
We now prove in Proposition~\ref{prop.near_oaps} that favorable propagation can be deduced from either the space $\mathcal{H}_1$ of matrices, as done in previous studies \cite[Sections 2.5.2 and 4.4.1]{bjornson2017massive}, or from the infinite dimensional space $\mathcal{H}_3$ of square-integrable functions. With this result, we derive in Corollary~\ref{cor.quality} a simple scheme to identify channels offering favorable propagation if the only  information available is the array response and rough knowledge of the angular power spectra.

\begin{proposition}
\label{prop.near_oaps}
Let $j\in\mathcal{U}$ and $l\in\mathcal{U}\backslash\{j\}$ be  users with angular power spectra $(\rho_j,\rho_l)\in\mathcal{K}\times\mathcal{K}$. Then 

\begin{align}
\label{eq.var}
0\le E[|\signal{h}_j^H\signal{h}_l|^2] = \innerprodh{\signal{R}_j}{\signal{R}_l}{\ha}=\innerprodh{\rho_j}{T^* T (\rho_l)}{\hc}
\end{align}
Furthermore,
\begin{align}
\label{eq.Chebyshev}
(\forall \epsilon>0)~\mathrm{Pr}(|\signal{h}_j^H\signal{h}_l|\ge \epsilon) \le \dfrac{1}{\epsilon^2} \innerprodh{\rho_j}{T^* T (\rho_l)}{\hc},
\end{align}
and the following are equivalent: (i) $\innerprodh{\signal{R}_j}{\signal{R}_l}{\ha}=0$, \linebreak[4] (ii) $\innerprodh{\rho_j}{T^* T(\rho_l)}{\hc}=0$, (iii) $\mathrm{Pr}(|\signal{h}_j^H\signal{h}_l| = 0) = 1$, and (iv) $\signal{R}_j\signal{R}_l=\signal{0}$.
\end{proposition} 

From a practical perspective, the inner product $\innerprodh{\rho_j}{T^* T (\rho_l)}{\hc}$ in \refeq{eq.var} and \refeq{eq.Chebyshev} explicitly reveals the individual influences of the signals impinging on the array (the angular power spectra $\rho_j$  and $\rho_l$) and the array response (the self-adjoint operator $T^*T$) on the interference pattern. This explicit separation of the contributions of impinging signals and the array response can be used to generalize in a unified way previous results such as the schedulers in \cite{mupp18,zhao2015,yin2014}. To be concrete, suppose that we know $\rho_l\in\mathcal{K}$, and let $\mathcal{Z}\subset\Omega$ be angles for which the function $T^*T(\rho_l)\in\mathcal{K}$ takes values close to zero. In this situation, we can schedule user $l$ together with a second user $j$ with $\mathrm{Supp}(\rho_j)\subset\mathcal{Z}$ to obtain interfering channels offering approximately favorable propagation (because $\innerprodh{\rho_j}{T^*T(\rho_l)}{\hc}$ will be small). Note that, to describe these schedulers, we have not simplified the model by considering discrete angular power spectra or particular array responses. Nevertheless, if the model is given, the proposed framework enables us to deduce, with simple and precise mathematical statements,  the potential problems caused by ignoring  practical aspects. For example, the operator $T^* T$ is typically a smoothing operator, so it is expected to increase the support of functions in $\mathcal{K}$. See, in particular, the integral form of $T^*T$ with the kernel $\kappa$ in  \refeq{eq.smoothing}, which is the standard form of a smoothing operator. As a result, \refeq{eq.var} shows that interference can be strong even if the angular power spectra of interfering users have disjoint support. The effects of  $T^*T$ on the interference pattern cannot be ignored. Proposition~\ref{prop.near_oaps} also indicates that peaks in $T^*T (1_\Omega)\in\mathcal{K}$ are angles for which the array is mostly susceptible to interference. 

In the above discussion, we have assumed knowledge of the interfering angular power spectra to identify channels offering approximately favorable propagation. The next result, which follows directly from Proposition~\ref{prop.nonnegative}, together with Proposition~\ref{prop.near_oaps} relaxes this requirement. 

\begin{Cor}
	\label{cor.quality}
	Let $\mathcal{C}=\{\rho\in\mathcal{K}~|~\mathrm{ess~sup}_{\theta\in\Omega}~ \rho(\theta) \le 1\}$ be the set of nonnegative functions essentially bounded by one\footnote{The assumption that the functions are bounded by the value one is just used for convenience. Any other value could be used, in which case an additional multiplying constant would appear in \refeq{eq.quality}.} and $\mathcal{M}\subset 2^\Omega$ be the set of Lebesgue-measurable subsets of $\Omega$. For two distinct users $(j,l)\in\mathcal{U}\times\mathcal{U}$, assume that their corresponding angular power spectra $(\rho_j,\rho_l)\in \mathcal{C}\times \mathcal{C}$ are \emph{unknown} functions with support in $(S_j, S_l)\in \mathcal{M}\times \mathcal{M}$. Then $$\innerprod{\rho_j}{T^*T(\rho_l)}_{\hc} \le Q(S_j,S_l,T),$$
	where 
	\begin{align}
\label{eq.quality} 
\begin{array}{rl}	Q:\mathcal{M}\times\mathcal{M}\times\mathcal{B}(\mathcal{H}_3,\mathcal{H}_2)&\to\real_+\\(\mathcal{X},\mathcal{Y},T)&\mapsto\innerprodh{1_{\mathcal{X}}}{T^*T(1_{\mathcal{Y}})}{\hc}
\end{array}
	\end{align}
 is called the \emph{quality function} of the array.
\end{Cor}

\subsection{Pilot decontamination for channel covariance estimation}
\label{sect:algorithms}

We now turn the attention to the algorithm that has the objective of estimating the covariance matrix of a desired user (supposed to be user $j=1$) in multiuser environments. To derive the algorithm, we  assume knowledge of the covariance matrix $\signal{R}$ in \refeq{eq.contamination}, the noise variance $\sigma^2$, and the mapping $T$. In particular, with a sufficient large design parameter $L\in\Natural$, the covariance matrix $\signal{R}$ can be realibly estimated with projection techniques applied to the sample covariance matrix $(1/L)\sum_{k=1}^L \signal{y}[k]\signal{y}[k]^H \approx \signal{R}$ (see \cite[Sect.~4.2]{miretti18} and the references therein). We further assume knowledge of two non-null Lebesgue-measurable sets $(\mathcal{M}_1,\mathcal{M}_\mathrm{int})\in 2^\Omega\times 2^\Omega$ such that (i) $\mathcal{M}_1 \supset \mathrm{Supp}(\rho_1)$, (ii) $\mathcal{M}_\mathrm{int}\supset \mathrm{Supp}(\sum_{j\in\mathcal{U}\backslash\{1\}}\rho_j)$,  and (iii) $Q(\mathcal{M}_1,\mathcal{M}_\mathrm{int}, T)$ [see \refeq{eq.quality}] is close to zero. Location-aided schedulers \cite{mupp18,zhao2015} can be used to guarantee the last condition. 

 Defining the ``denoised'' covariance matrix $\signal{R}_\mathrm{d}\in\mathbb{C}^{N\times N}$ by $\signal{R}_\mathrm{d}:= \signal{R}-\sigma^2\signal{I}=\signal{R}_1+\sum_{j\in\mathcal{U}\backslash\{1\}}\signal{R}_j,$ we can informally describe the proposed algorithm with three simple steps:
\begin{enumerate}
	\item Obtain an estimate $\tilde{\rho}\in\mathcal{H}_3$ of  $\rho_\mathrm{d}:=\sum_{j\in\mathcal{U}} \rho_j\in\mathcal{K}$ from $\signal{R}_\mathrm{d}$, or, equivalently, from $\signal{r}_\mathrm{d}:=T_\mathrm{vec}(\signal{R}_d)$.
	\item Use knowledge of $\mathcal{M}_1\supset \mathrm{Supp}(\rho_1)$ to compute an estimate $\tilde{\rho}_1$ of  $\rho_1$ from $\tilde{\rho}$ obtained in Step 1.
	\item Use $\tilde{\rho}_1$ in Step 2 and knowledge of $T\in\mathcal{B}(\mathcal{H}_3, \mathcal{H}_2)$ to compute the estimate $T_\mathrm{vec}(\tilde{\signal{R}}_1)$ of $T_\mathrm{vec}(\signal{R}_1)$. 
\end{enumerate}
Later, in Proposition~\ref{prop.summary}, we show that all steps can be implemented with a simple matrix-vector multiplication, but first we need to detail the operations in each step.

({\it\bf Step 1:}) Recalling that the operator $T$ in \refeq{eq.bounded_op} is linear, we have $\sum_{j\in\mathcal{U}}T (\rho_j) = T\left(\sum_{j\in\mathcal{U}}\rho_j\right)=T(\rho_\mathrm{d})  =T_\mathrm{vec}(\signal{R}_\mathrm{d})=:\signal{r}_\mathrm{d}$. The equality $T(\rho_\mathrm{d})=\signal{r}_\mathrm{d}$ shows that estimating ${\rho}_\mathrm{d}$ given $\signal{r}_\mathrm{d}$ is an ill-posed problem similar to that described in \cite{miretti18,miretti18SPAWC,renato18error}, so it can be addressed with any algorithm proposed in those studies. In particular, for computational simplicity and brevity, we focus on the scheme in \cite[Sect.~3.1]{miretti18}, which estimates $\rho_\mathrm{d}$ by solving 
\begin{align*}
\text{minimize}_{\rho\in\{h\in\mathcal{H}_3~|~T(h)=\signal{r}_d\}\neq\emptyset} \|\rho\|_\hc, 
\end{align*}
where $\|\cdot\|_\hc:=\sqrt{\innerprodh{\cdot}{\cdot}{\hc}}$ is the norm induced by the inner product. The above problem has a unique solution given by  \cite[Ch.~6]{luen} \linebreak[4] $\tilde{\rho}:= T^\dagger (\signal{r}_\mathrm{d}) = \sum_{n=1}^{2N^2} \alpha_n g_n \in\mathcal{H}_3$,
where $T^\dagger:\mathcal{H}_2\to \mathcal{H}_3$ denotes the pseudo-inverse of the operator $T$,  $(g_n)_{n\in\{1,\ldots,2N^2\}}$ are the functions in \refeq{eq.innerprod},  $\signal{\alpha}=[\alpha_1,\ldots,\alpha_{2N^2}]^t$ is any solution to $\signal{G}\signal{\alpha}=\signal{r}_\mathrm{d}$, and the component $\signal{G}_{n,m}$ of the $n$th row and $m$th column of the Gramian matrix $\signal{G}\in\real^{2N^2 \times 2N^2}$ is given by 
\begin{align}
\label{eq.G}
\signal{G}_{n,m}=\innerprodh{g_n}{g_m}{\hc}.
\end{align}

{\it\bf (Step 2:)} Since $Q(\mathcal{M}_1, \mathcal{M}_\mathrm{int}, T)$ is assumed to be sufficiently close to zero, the only signal in $\tilde{\rho}\approx\sum_{j\in\mathcal{U}}\rho_j$ with significant power at angles in the set $\mathcal{M}_1$ is expected to be the signal of user $j=1$. Therefore, we use $\tilde{\rho}_1 = P_{C_1}(\tilde{\rho}) \in\mathcal{H}_3$
as the estimate of $\rho_1$, where  $P_{C_1}:\mathcal{H}_3\to C_1:\rho\mapsto 1_{\mathcal{M}_1}\cdot{\rho} $ denotes the orthogonal projection onto the closed subspace $C_1:=\overline{\{\rho \in \mathcal{H}_3~|~(\forall \theta \notin \mathcal{M}_1) \rho(\theta)=0\}}$ of functions with support in $\mathcal{M}_1$ (the overline operator denotes the closure of a set). In this step, since $P_{C_1}$ is a projection onto a closed subspace, we have \cite[Corollary 3.24]{baus17} $$\|\rho_1-\tilde{\rho}_1\|^2_\hc = \|\rho_1-\tilde{\rho}\|^2_\hc-\|\tilde{\rho}-\tilde{\rho}_1\|^2_\hc.$$

{\it\bf (Step 3:)}
With $\tilde{\rho}_1$ obtained in Step 2, we compute the estimate $\tilde{\signal{r}}_1:=T_\mathrm{vec}(\tilde{\signal{R}}_1)$ of $T_\mathrm{vec}(\signal{R}_1)$ using \refeq{eq.bounded_op}, or, more precisely,  
\begin{align}
\label{eq.est_cov}
\tilde{\signal{r}}_1=T(\tilde{\rho}_1).
\end{align}
Note that \refeq{eq.est_cov} entails the evaluation of integrals whenever a new estimate of $\tilde{\rho}_1$ is available, even if the information about the support $\mathrm{Supp}(\rho_1)$ is fixed. This operation can be burdensome if the integrals are not easy to evaluate. However, the next proposition shows that all steps of the proposed algorithm can be combined in a single operation involving only one simple matrix-vector multiplication. In this equivalent form of the algorithm, integrals are evaluated to construct a matrix only if  information about $\mathrm{Supp}(\rho_1)$ is updated.

\begin{proposition}
	\label{prop.summary}
	Let $\signal{G}_{C_1}\in\real^{2N^2\times 2N^2}$ be the matrix with its $n$th row and $m$th column given by $(\signal{G}_{C_1})_{n,m} = \innerprodh{g_n}{P_{C_1} g_m}{\hc}$, and define $\signal{A}:=\signal{G}_{C_1}\signal{G}^\dagger\in\real^{2N^2\times 2N^2}$, where $\signal{G}^\dagger$ is the Moore-Penrose pseudo-inverse of the matrix $\signal{G}$ with entries given in \refeq{eq.G}. With $\tilde{\rho}_1$ estimated with the approaches in Steps 1 and 2, we have
	\begin{align}
	\label{eq.linear_proc}
	T_\mathrm{vec}(\tilde{\signal{R}}_1)=\tilde{\signal{r}}_\mathrm{1} = T(\tilde{\rho}_1) = \signal{A}\signal{r}_\mathrm{d}.
	\end{align}
\end{proposition}

\section{Simulations}
\vspace{-.2cm}
\label{sect.simulations}
For simplicity, we consider a system where a base station is equipped with the array in Example~\ref{example.ULA} with $f=2.11$ GHz, $c=3\cdot 10^8$ m/s, and $d=c/(2f)$. 
As prior knowledge given to the proposed algorithm, we assume that the unknown power spectrum $\rho_1$ of the desired user has most of its power in the interval $\mathcal{M}_1=[0.3, 1.2]$, and the combined angular power spectrum $\rho_\mathrm{int}=\sum_{j\in\mathcal{U}\backslash\{1\}}\rho_j$ of the interfering users has most of its power in the interval $\mathcal{M}_\mathrm{int}=[-1, 0]$. As in \cite[Sect.~5]{miretti18}, in each run of the simulation we use $\rho_1:\Omega\to\real_+:\theta\mapsto\sum_{k=1}^Q \alpha_k h_k(\theta)$, where $Q$ is uniformly drawn from $\{1,2,3,4,5\}$; $h_k:\Omega\to\real_+:\theta\mapsto ({1}/{\sqrt{2\pi\Delta_k^2}})\exp\left({-{(\theta-\phi_k)^2}/{(2\Delta_k^2)}}\right)$;
$\phi_k$, the main arriving angle of the $k$th path, is uniformly drawn from $[0.5, 1]$; and $\alpha_k$ is uniformly drawn from $[0,~1]$, and it is further normalized to satisfy $\sum_{k=1}^Q\alpha_k=1$. The combined angular power spectrum $\rho_\mathrm{int}$ of the interfering users is obtained similarly, with the difference that we draw the main arriving angles uniformly at random from $[-1,-0.5]$.

The performance of the proposed algorithm in \refeq{eq.linear_proc} is evaluated by considering both (i) perfect knowledge of the covariance matrix $\signal{R}_\mathrm{d}$ (Proposed-perfect) and (ii) its estimate \linebreak[4] $\widetilde{\signal{R}}_\mathrm{d} = P_\mathcal{T}((1/L)\sum_{k=1}^{L} \signal{y}[k]\signal{y}[k]^H -\sigma^2\signal{I} )$ (Proposed-estimate), where $L=1,000$ and  $P_\mathcal{T}:\mathbb{C}^{N\times N}\to\mathcal{T}$ is the projection onto the set $\mathcal{T}$ of Toeplitz and Hermitian positive semidefinite matrices  (see \cite[Sect.~4.2]{miretti18}) w.r.t. the complex Hilbert space $(\mathbb{C}^{N\times N},\innerprod{\signal{A}}{\signal{B}}=\mathrm{tr}(\signal{B}^H \signal{A}))$. These two variants of the proposed algorithm are compared with a baseline scheme that estimates directly the desired channel covariance matrix in an interference-free and noiseless system according to $\widetilde{\signal{R}}_1 = P_\mathcal{T}( (1/L) \sum_{k=1}^{L} \signal{h}_1[k]\signal{h}_1[k]^H)$. We use $(\forall j\in\mathcal{U})(\forall k\in\Natural)s_j[k]=1$ as pilots in \refeq{eq.signal}, and the noise samples are drawn from the distribution $\mathcal{N}_\mathbb{C}(\signal{0},\sigma^2\signal{I})$ with $\sigma^2 = 0.1$. In the proposed scheme, the integrals used to construct the matrix $\signal{G}$ in \refeq{eq.linear_proc} are computed with the closed-form expressions available in \cite{miretti18,renato18error}, whereas the integrals used to construct $\signal{G}_{C_1}$ and the covariance matrices are computed numerically. 
 The figure of merit is the (normalized) mean square error (MSE), which we define by $E\left[{\|\tilde{\signal{R}}_1 - \signal{R}_1\|_F^2}/{\|\signal{R}_1\|_F^2}\right]$, where $\|\cdot\|_\mathrm{F}$ is the Frobenius norm, and the expectation is approximated with the empirical average of 1,000 runs of the simulation. 
 
\begin{figure}
\begin{center}
\includegraphics[width=\columnwidth]{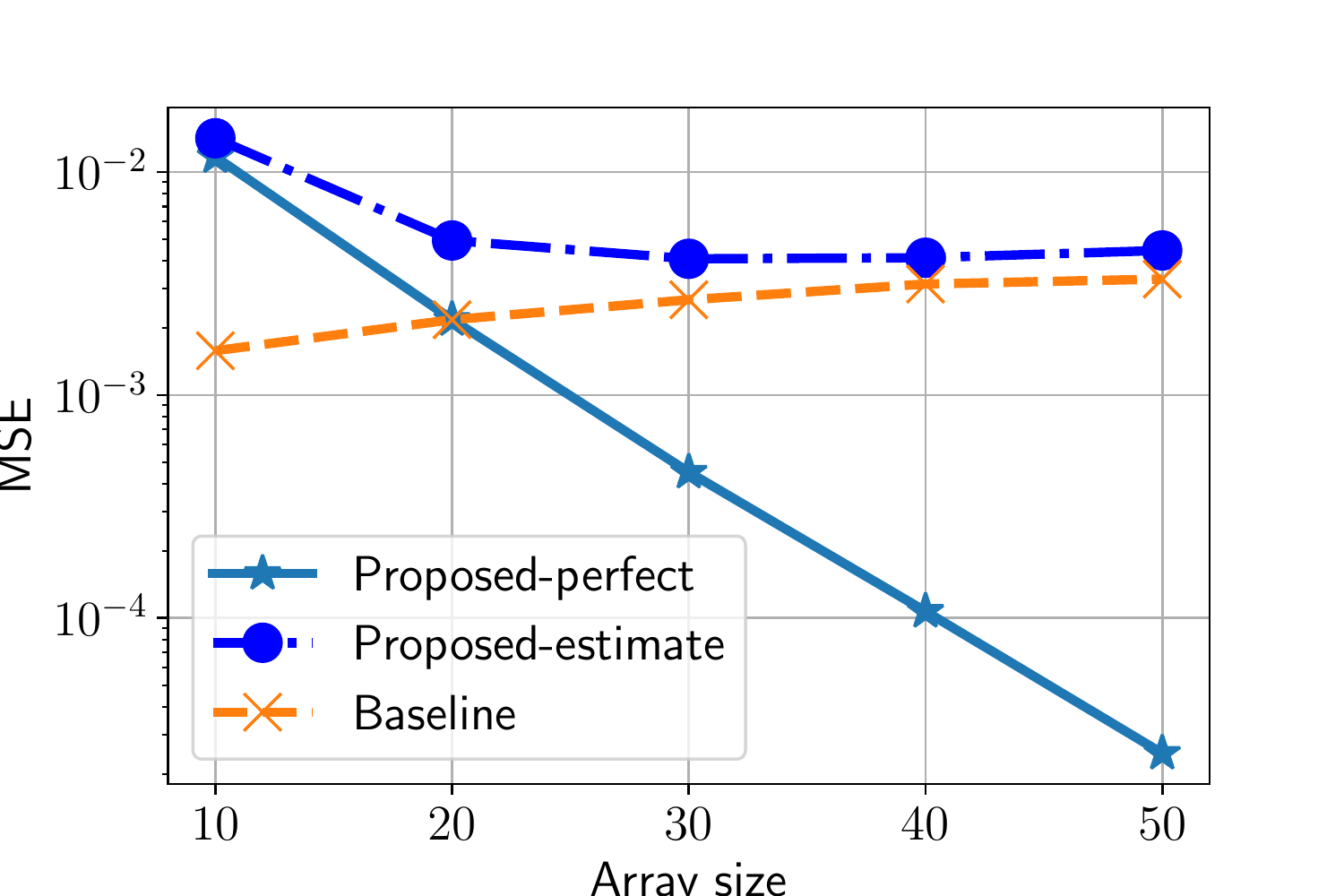}
\caption{Mean square error of the desired channel covariance matrix as a function of the number $N$ of antennas.}
\label{fig.bad_separation}
\end{center}
\vspace{-.8cm}
\end{figure}

 Fig.~\ref{fig.bad_separation} shows the MSE performance as a function of the number $N$ of antennas. The performance of the baseline scheme degrades with increasing $N$ because the sample estimate of $\signal{R}$ also degrades if the number $L$ of samples is fixed. With $N$ small, the performance of Proposed-perfect and Proposed-estimate are similar, but worse than that obtained with the baseline scheme, which indicates that the practical algorithm Proposed-estimate is limited by the angular resolution of the array. This fact can also be verified by plotting the quality function $Q(\mathcal{M}_1,~\mathcal{M}_\mathrm{int}, T)$ in \refeq{eq.quality} for the different array sizes, but we do not show this figure because of the space limitation. With large arrays, Proposed-estimate has performance similar to the baseline scheme, whereas Proposed-perfect has the best performance because it uses perfect knowledge of the matrix $\signal{R}$, and the array offers enough angular resolution. These performance relations indicate that, with $N$ sufficiently large, Proposed-estimate is mostly limited by the accuracy of the estimate of $\signal{R}$, and not by interference.

\section{Summary and conclusions}
We have shown that common simplifications used to study interference in massive MIMO systems (e.g., discrete angular power spectra, infinitely large arrays, etc.) can be avoided by using standard results in functional and convex analysis. In particular, we showed that the effects of signals impinging on an array and the array response on the interference pattern can be decoupled in an infinite-dimensional Hilbert space. This natural decomposition enabled us to devise a simple algorithm to estimate the channel covariance matrix of a given user in multi-user systems. In some scenarios, the algorithm shows performance similar to that obtained with direct sample-based estimation techniques operating in interference-free and noiseless systems. Its performance is mainly limited by the availability of accurate estimates of the covariance matrix of the input signal.  

\bibliographystyle{IEEEtran}
\bibliography{IEEEabrv,references}

\end{document}